\documentclass[]{spie}

\usepackage{amsmath,amsfonts,amssymb}
\usepackage{graphicx}
\usepackage[colorlinks=true, allcolors=blue]{hyperref}
\usepackage{subfig}
\usepackage{caption}
\usepackage{subcaption}

\title{The beam test facility at Jefferson Lab for the precise energy measurement of future calorimeter concepts}

\author[a]{Vladimir V. Berdnikov}
\author[a]{Alexander Somov}
\affil[a]{Thomas Jefferson National Accelerator Facility, Newport News, Virginia 23606, USA}

\authorinfo{Further author information: (Send correspondence to Vladimir V. Berdnikov)\\Vladimir V. Berdnikov: E-mail: berdnik@jlab.org, Telephone: +1-757-269-6928\\ Alexander Somov: E-mail: somov@jlab.org, Telephone: +1-757-269-5553}

\begin{document} 
\maketitle

\begin{abstract}
In experimental nuclear physics (NP), high-precision electromagnetic calorimetry typically requires a good energy resolution and linear photosensor response on the level of (1–2)$\%$ over a full dynamic range of the detector. The beam of secondary leptons at the Jefferson Lab experimental complex, provided by the Hall D pair spectrometer (PS), is an optimal facility for studies aiming to understand the impact of light collection and signal processing on calorimetry energy resolution under real experimental conditions. The light emission and collection processes depend on radiator component quality and type, while signal processing depends on photosensor type and front-end electronics design. Various calorimeter tower components and detector assemblies for current and future NP experiments were tested within the lepton momentum range of (3–6) GeV/c using PS. The energy resolution of the PS itself is estimated to be better than $\approx$0.6$\%$.

\end{abstract}
 

\section{INTRODUCTION}

In nuclear physics, high-precision electromagnetic calorimetry requires the use of specific homogeneous radiator materials. One of the most prominent examples is lead tungstate crystals (PbWO$_4$). These crystals are considered long lead-time items and can be produced by only one or two vendors worldwide, none of which are located in the United States. The acquisition process demands detailed characterization and strict quality control~\cite{Horn:2019beh, Houzvicka:2022pny}. Different crystal types were characterized using a beam of secondary leptons provided by the Hall D experimental complex at Jefferson Lab (JLab). 
Recently, there has been significant interest in developing alternative radiator materials that are radiation-hard and suitable for mass production. SciGlass is one attractive and cost-effective material, though it is still in the R$\&$D phase and requires extensive testing~\cite{Horn:2025hnz}. The new material was tested using the Pair Spectrometer test beam area. 

The PS setup was employed to evaluate various detector prototypes, which played a crucial role in the development of large-scale detectors such as the CCAl, NPS, and ECAL~\cite{Asaturyan:2021ese}. Our investigations focused on the linearity of the detector response and front-end electronics, as well as the optimization of the supplied voltage parameters. Accurate energy reconstruction depends on a clear understanding of detector linearity across the full operational dynamic range. In addition, we optimized the designs of calorimeter modules and their mechanical structures, aiming to determine how different configurations impact energy resolution.

In future collider experiments at the Electron-Ion Collider (EIC) detector the application of a PMT is almost impossible due to the presence of a strong magnetic field. Several lead tungstate calorimeter prototypes were built using newly available SiPMs with a small pixel pitch, optimized for large dynamic range and linearity. However, deviations from linearity in typical silicon photomultipliers (SiPMs) currently available on the market can already appear at energies of a few GeV. The impact of SiPM saturation effects on energy resolution under realistic experimental conditions was evaluated and discussed~\cite{Ameli:2022gvw,Philip:2023jgx,Philip:2024opz,berdnikov2023poster,berdnikov2024poster}. 
Overall, since 2018 more than twenty various calorimeter concepts have been installed and tested with the Pair spectrometer in Hall D.

\section{Pair Spectrometer in Hall D}

\begin{figure}[t]
\centering
\subfloat[Schematic of the Pair Spectrometer area.]{
\includegraphics[width=.35\textwidth]{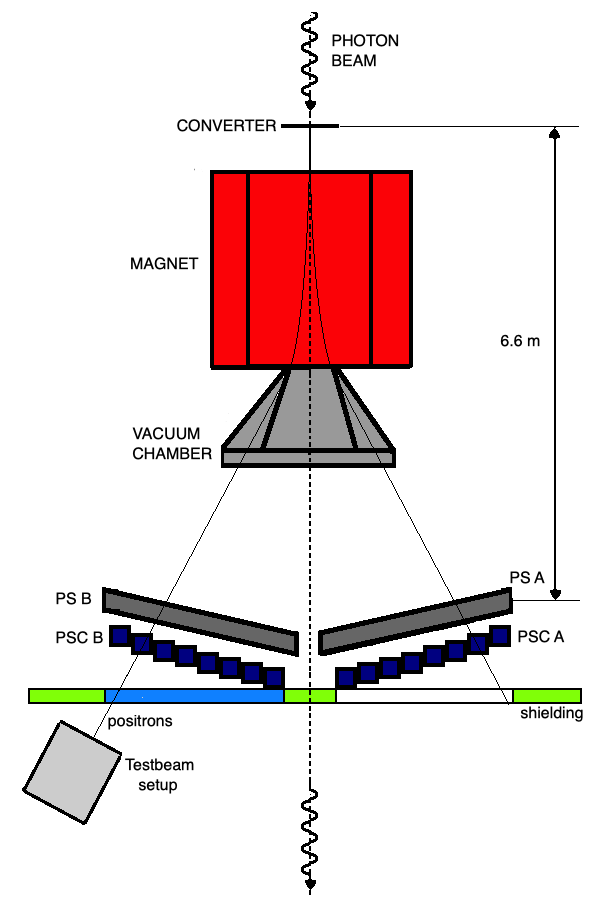}
}
\hspace{0.5in}
\subfloat[Overview of the Pair Spectrometer area.]{
\includegraphics[width=.4\textwidth]{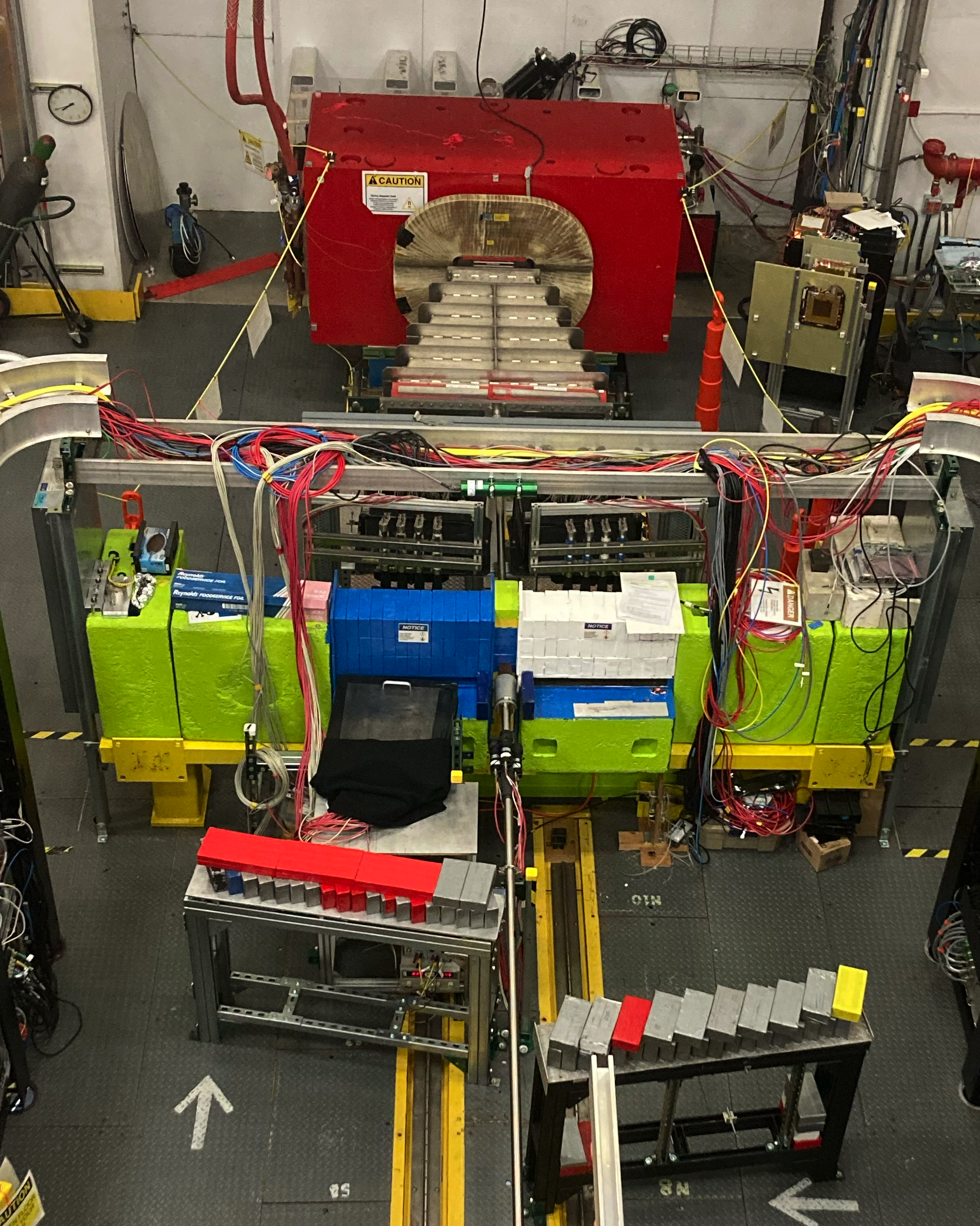}
}
\caption{Pair Spectrometer in Hall D.}
\label{fig1}
\end{figure}

The performance of the calorimeter prototype was studied using secondary electrons generated by the Hall D Pair Spectrometer (PS)~\cite{ps,ps1}. A schematic view of the Pair Spectrometer setup is presented in Fig.~\ref{fig1}a. In this system, electron-positron pairs are produced when incident beam photons interact with a 750 $\mu$m Beryllium converter. The resulting leptons are then directed into a 1.5 Tesla dipole magnet, which bends their trajectories according to their momenta. The spectrometer is equipped with two symmetric detector arms, located on either side of the photon beam axis. Each arm consists of two layers of scintillator counters: eight coarse counters with larger active areas, primarily used for triggering, and 145 high-granularity counters, which offer fine spatial resolution. The high-granularity hodoscope enables accurate momentum reconstruction by correlating the hit position of a lepton with its deflection angle in the magnetic field. This design allows the spectrometer to cover a broad range of electron and positron momenta, from approximately 3.0 GeV/c up to 6 GeV/c. The energy resolution of the spectrometer, crucial for precise energy tagging, is estimated to be better than 0.6$\%$, making it suitable for high-precision calorimeter calibration and performance studies. 

The calorimeter prototypes were positioned behind the Pair Spectrometer, as illustrated in Fig.~\ref{fig1}. The detector assemblies were carefully aligned using a laser survey and alignment technique system to ensure accurate positioning up to the precision of 200 $\mu$m relative to the beamline and the PS hodoscope. Special attention was given to aligning the calorimeter tower front face with the nominal beam path.

High-voltage power for individual channels of the calorimeter prototypes was supplied by a CAEN A1535SN module, known for its stability and compatibility with precision detector applications. The analog signals produced by the calorimeter were digitized using a fast analog-to-digital converter (ADC) module operated at a sampling rate of 250 MHz. fADC module was installed in the existing VXS crate, which is part of the infrastructure used for the Hall D main experimental program. The data from the prototype was acquired in parallel with Hall D beam operations, triggered by events identified by the Pair Spectrometer.

\begin{figure}[t]
\centering
\subfloat[Calorimeter radiator tests stand assembly for 12 modules.]{
\includegraphics[angle=270, width=.35\textwidth]{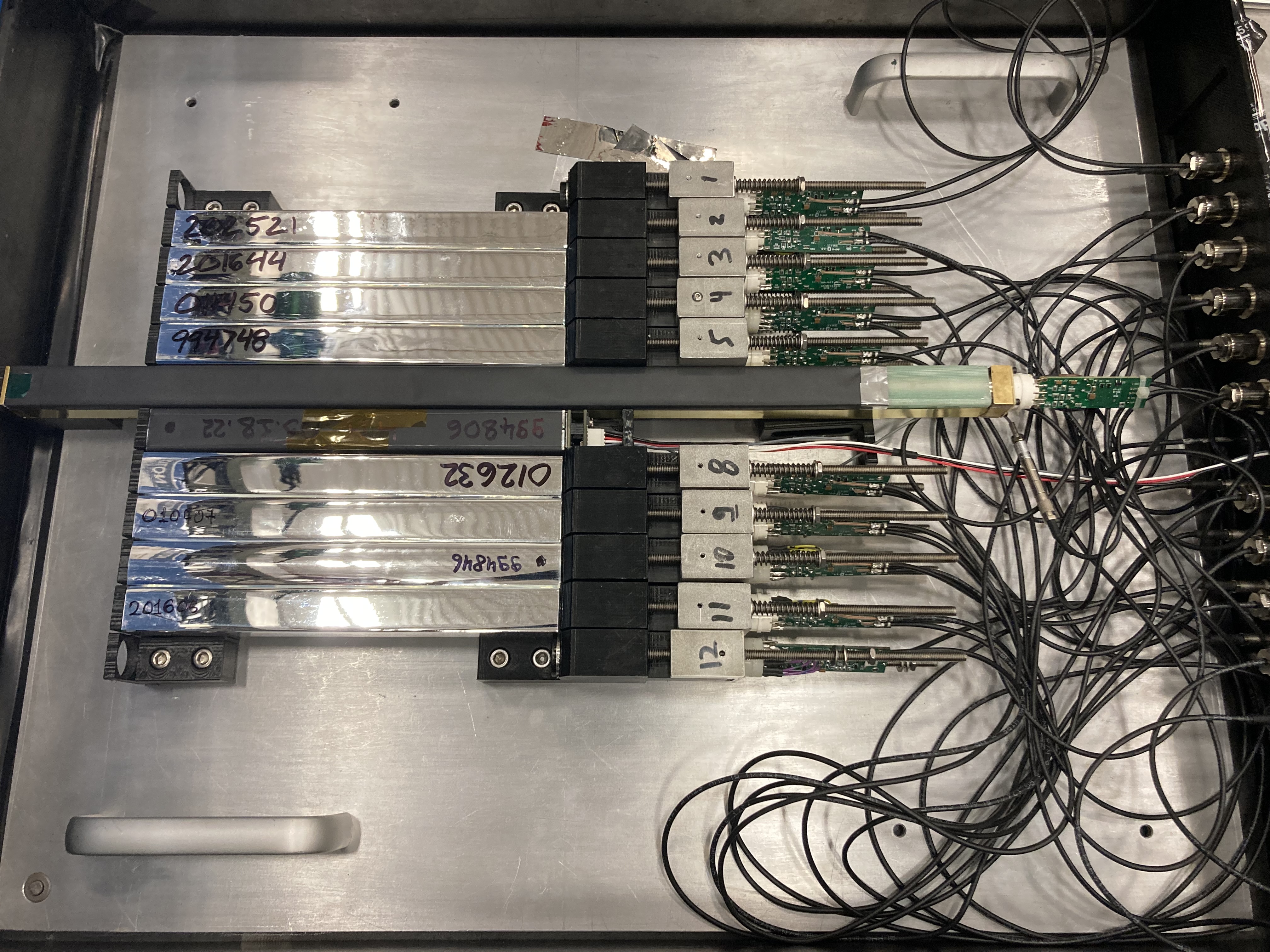}
}
\hspace{0.5in}
\subfloat[Calorimeter radiator test stand installed downstream of the beam from the PS.]{
\includegraphics[angle=270, width=.35\textwidth]{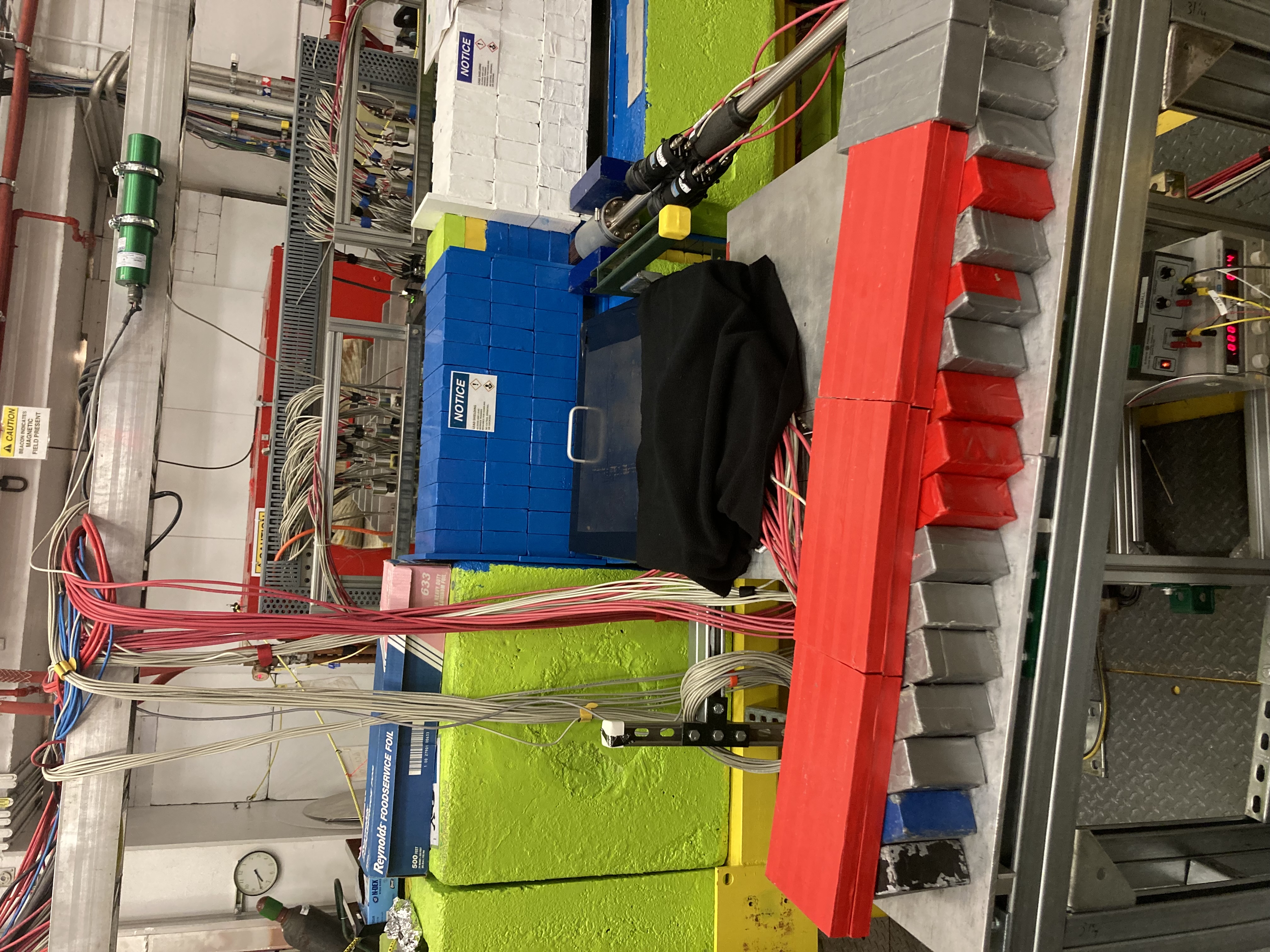}
}
\caption{Pair Spectrometer test stand for calorimeter radiators.}
\label{fig2}
\end{figure}

\section{Individual calorimeter towers characterization}
\label{sec:sections}
\subsection{Testing PbWO$_4$ crystals with PMT readout}
The first calorimeter tests were performed in 2018 during construction of the Compton calorimeter for the PrimEx-eta experiment in Hall D. More than 100 PbWO$_4$ crystals were checked using the test setup shown in Fig.~\ref{fig2}a.
The mechanical structure of the test stand accommodated twelve calorimeter towers mounted on an aluminum frame. The frame was placed inside a light-tight stainless steel box. Individual crystals 20.05$\times$20.05$\times$200.0 mm$^3$ were wrapped with 3M ESR 60 $\mu$m polymer film, which provides 98.5$\%$ reflectivity across the visible spectrum. The crystals were coupled with R4125-01 Hamamatsu PMTs using optical grease. Each crystal was placed in a special slot between two 3D-printed mounting components. 
Optical coupling pressure between the back face of the crystal and the PMT was provided by a spring-loaded aluminum mechanism mounted to the divider.
The PMT was powered and read out using an HV divider with an integrated preamplifier designed at Jefferson Lab. High voltage and signal cables were connected via SHV and LEMO connectors installed on the back patch panel of the test stand frame. The front face of the frame was covered with Tedlar film to prevent light leakage and minimize material in the particle path.
High voltages of 1 kV were set for all PMTs to ensure that the signal amplitudes did not saturate the ADC range, which maximum range corresponded to 2 V. Signal pulse information was read out from the ADC using two modes: signal waveforms without sparsification (no thresholds applied) and processed pulse information, such as pulse integral and peak amplitude. 

\subsection{Testing PbWO$_4$ crystals with SiPM readout}

Originally, the test stand was designed to characterize lead tungstate crystals with PMTs only, but it was later adopted for additional studies. 
The SiPM-based module used a lead tungstate crystal coupled to a 4×4 array of Hamamatsu S14160-3010PS SiPMs mounted on a printed circuit board (PCB), attached to a 3D-printed Acrylonitrile Butadiene Styrene (ABS) holder at the rear. A pair of brass-strapped flanges held the assembly. The front-end electronics consisted of two stacked PCBs: the first hosted the surface-mount SiPMs, while the second contained the preamplifier and connectors. All 16 SiPMs were powered by a common bias voltage, with an expected gain variation of 1–2$\%$ between devices. The temperature variation was compensated using a thermistor. Signals were digitized using fADC module positioned in VXS crate.

\begin{figure}[t]
\centering
\subfloat[Signal pulse integral as a function of the PS tile.]{
\includegraphics[width=.4\textwidth]{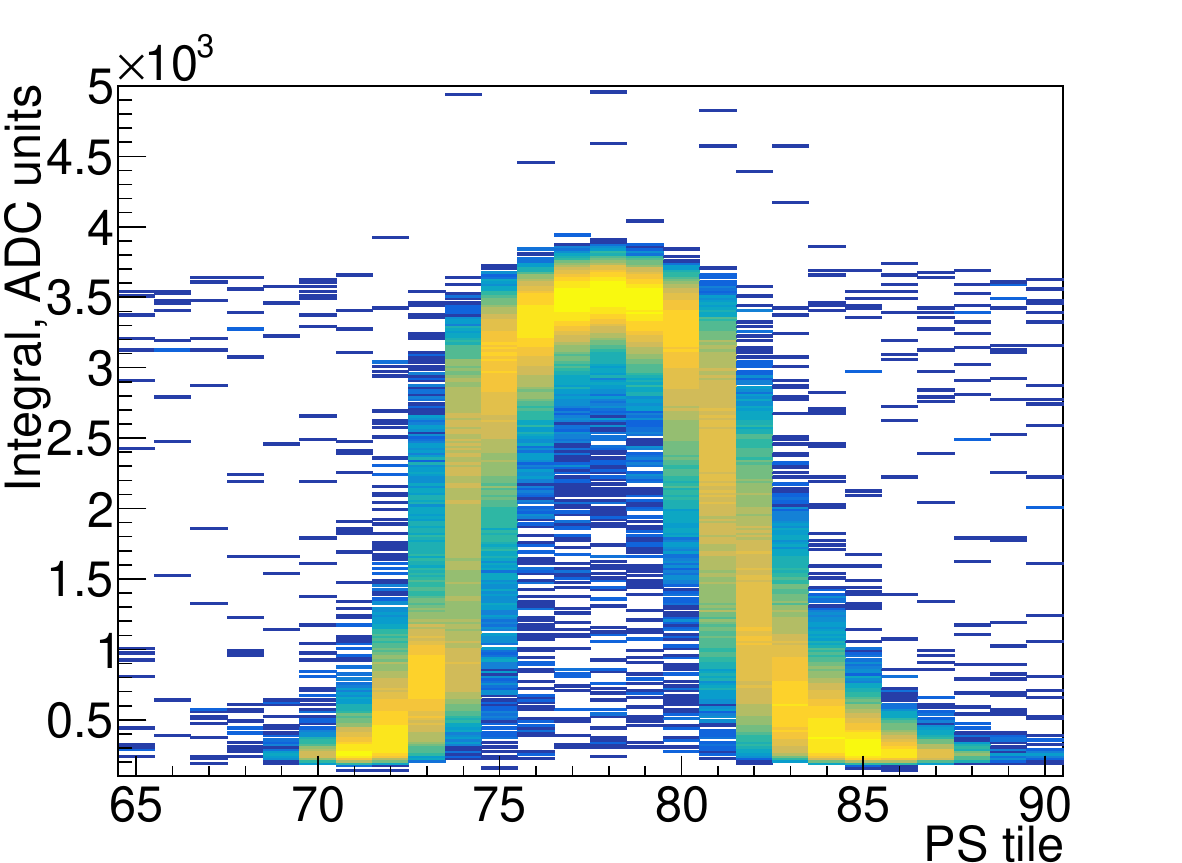}
}
\subfloat[Signal pulse integral for the lepton energy of 4.7 GeV.]{
\includegraphics[width=.58\textwidth]{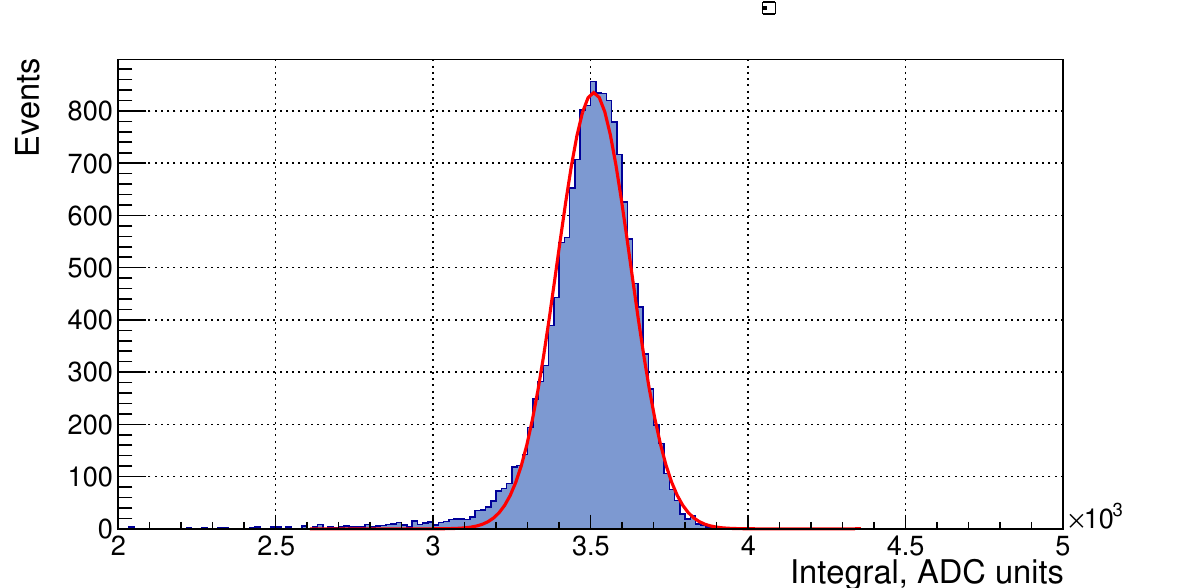}
}
\caption{Energy distributions for the PbWO$_4$ modules with the SiPM readout.}
\label{fig4}
\end{figure}

Fig.~\ref{fig4}a shows the pulse integral response, in units of ADC counts, of individual calorimeter towers as a function of the PS tile number. Each PS tile corresponds to a specific lepton energy. A lepton passing through the center of a crystal corresponds to the PS tile located at the center of the plateau, with an energy of approximately 4.7 GeV. The energy distribution of the crystal is shown in Fig.~\ref{fig4}b. Superimposed on the plot is the result of a fit to the Crystal Ball function.

The setup was later used to study the properties of a scintillating glass (SciGlass) radiator. SciGlass modules, each measuring $20\times 20\times 400$ 
mm$^3$, were subsequently installed on the PS test stand. Each module was wrapped in ESR reflective foil and coupled to a PMT housed in a G-10 enclosure. Two flanges at the crystal and housing ends were secured with a 25$\mu$m brass strap to ensure mechanical stability. The same high-voltage power supply and front-end electronics were used.
   
\section{Calorimeter prototypes characterization}

\begin{figure*}[b]
\centering
  \begin{minipage}[b]{0.25\textwidth}
    \centering
    \includegraphics[angle=270,width=\linewidth]{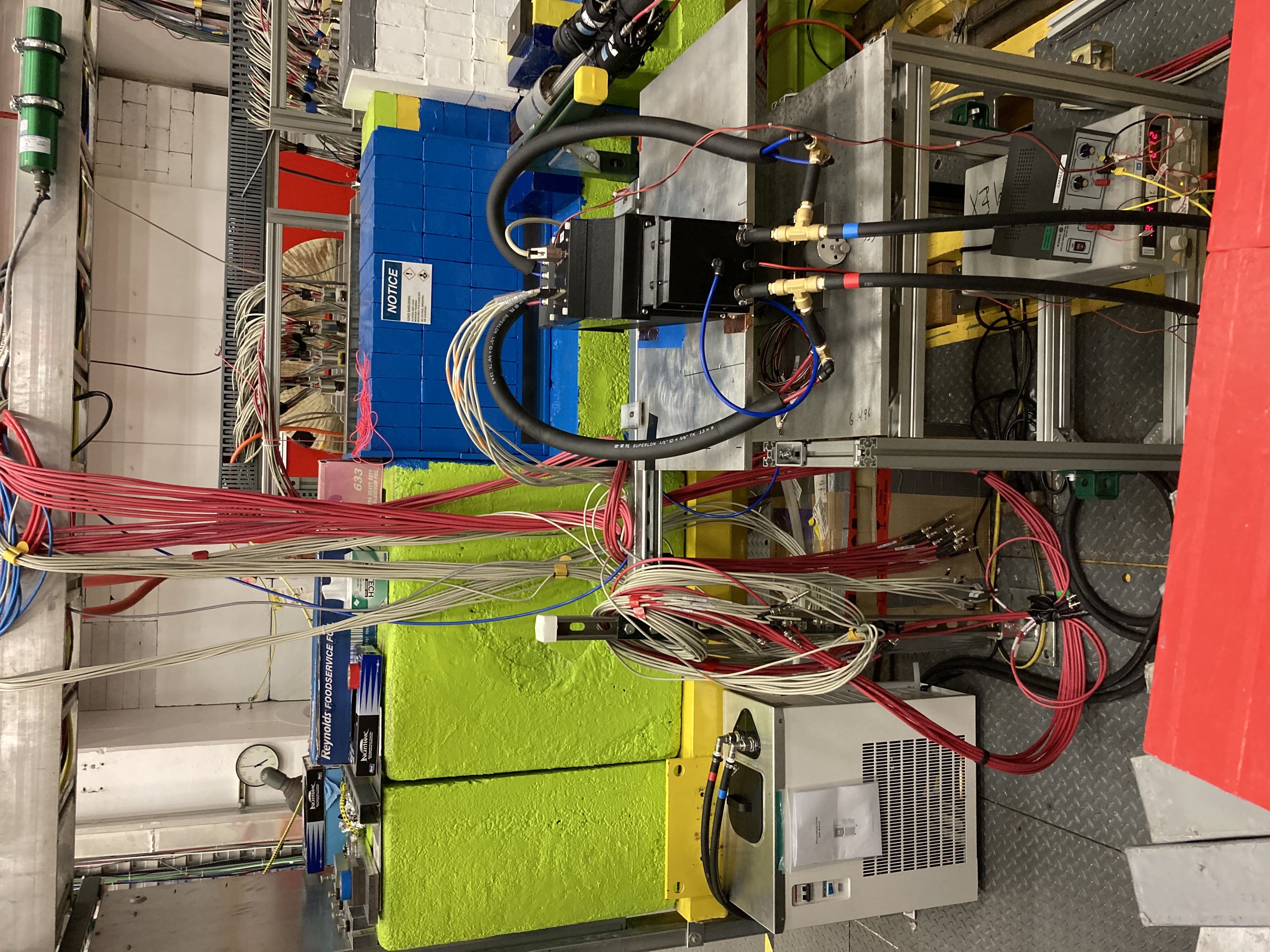}
    \captionof{figure}{JLab prototype with SiPM readout.}
    \label{fig5}
  \end{minipage} \hfill
  \begin{minipage}[b]{0.25\textwidth}
    \centering
    \includegraphics[angle=270,width=\linewidth]{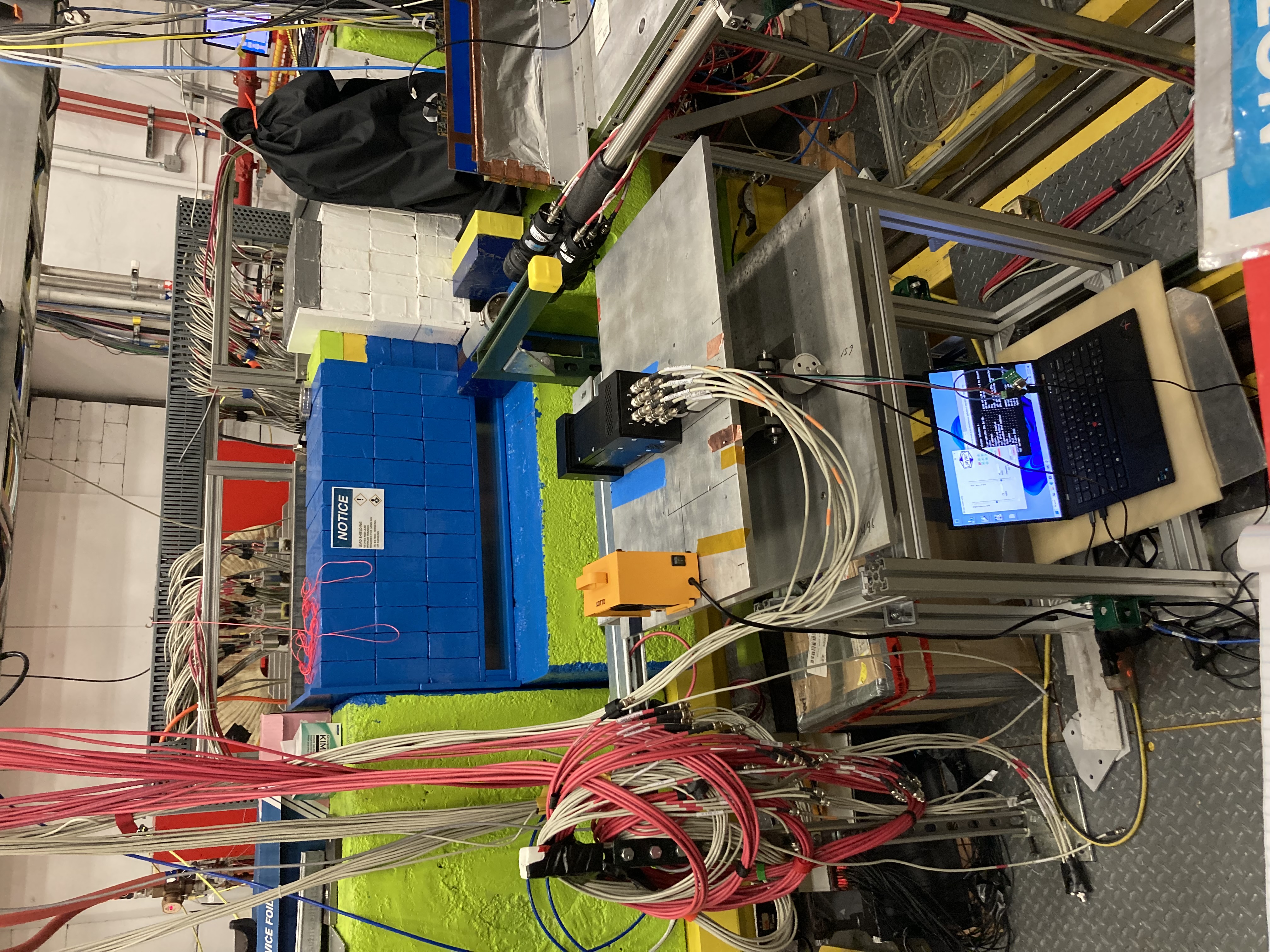}
    \captionof{figure}{CRYTUR prototype with SiPM readout.}
    \label{fig6}
  \end{minipage} \hfill
  \begin{minipage}[b]{0.4\textwidth}
    \centering
    \includegraphics[width=\linewidth]{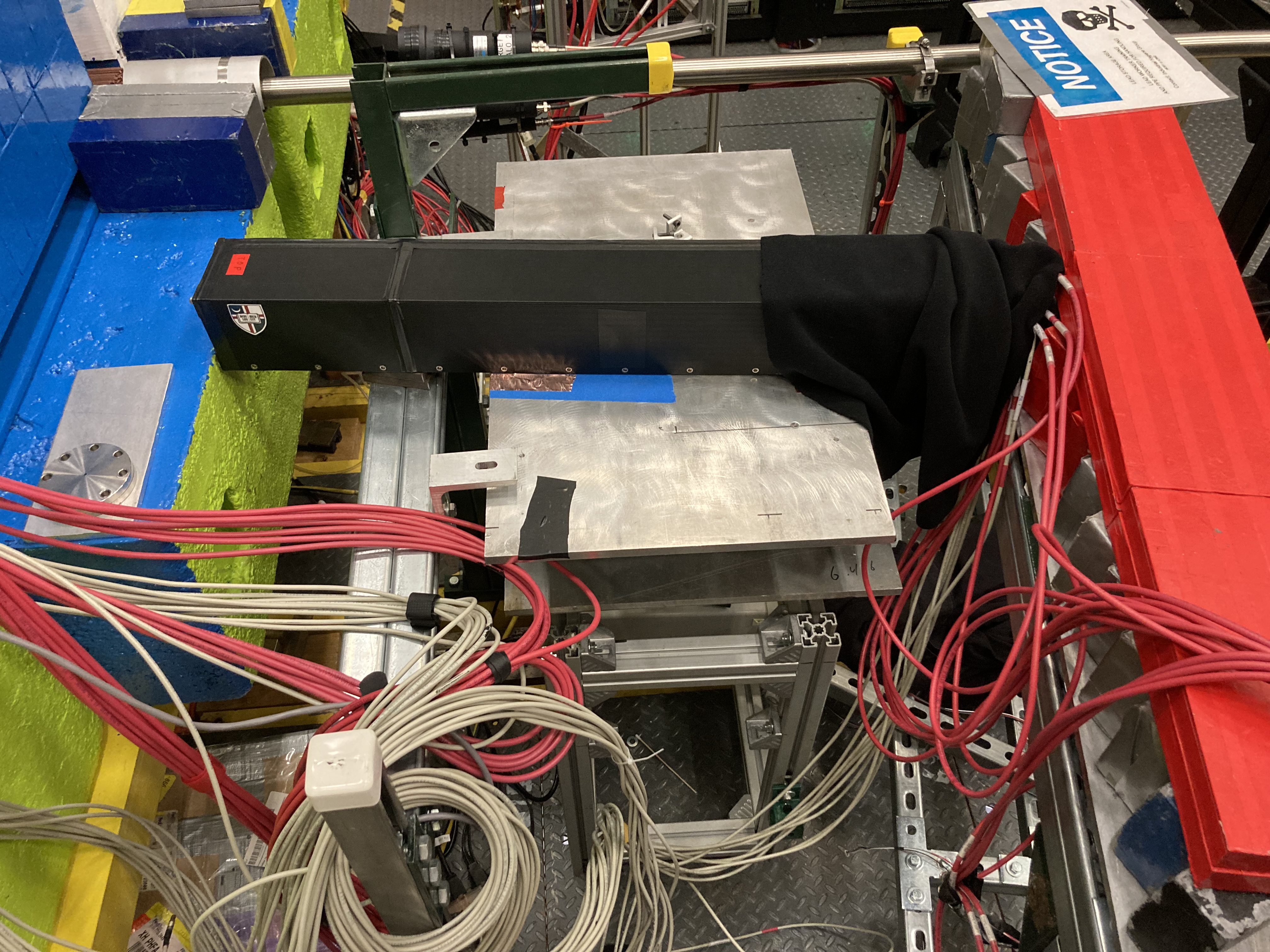}
    \captionof{figure}{SciGlass prototype with PMT readout.}
    \label{fig7}
  \end{minipage}
\end{figure*}

Several calorimeter prototypes with different readout types, radiators, and photosensors were tested using the Pair Spectrometer. Typically, they represented a calorimeter tower array with dimensions of 3×3 or larger, placed in a light-tight frame, as shown in Fig.~\ref{fig5}–\ref{fig7}. Detectors were mounted on a movable table to ensure precise alignment so that the leptons hit the central row of towers perpendicularly to the radiator’s front face. In the case of SiPM readout, water or air cooling was used to stabilize the temperature. 

\begin{figure}[t]
\centering
\subfloat[Calorimeter pulse integrals as function of the PS tile. ]{
\includegraphics[width=.5\textwidth]{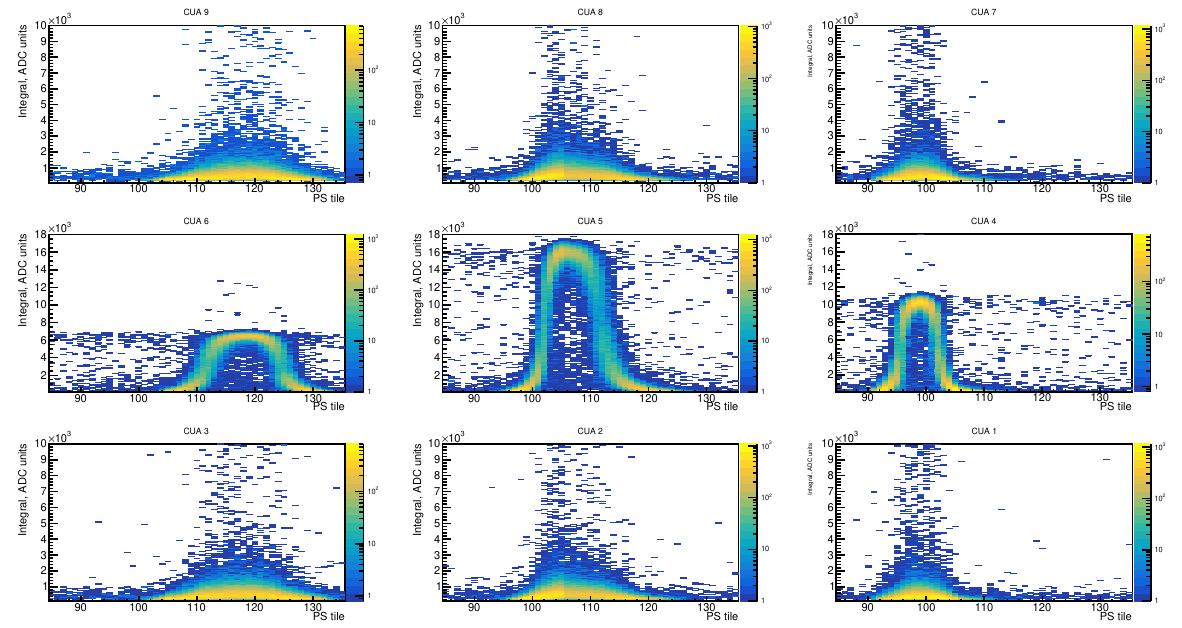}
}
\subfloat[Energy distribution deposited by 5 GeV/c$^2$ positrons in a single module (green) and the sum of 9 modules (red).]{
\includegraphics[width=.47\textwidth]{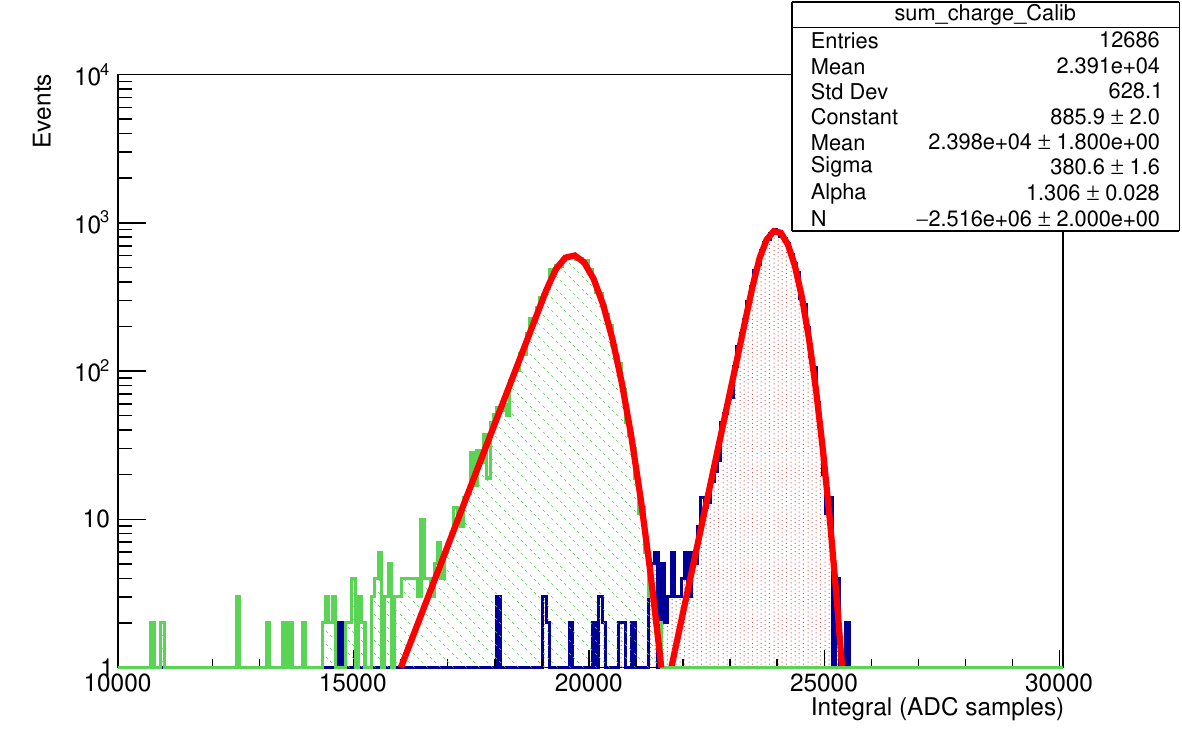}
}
\caption{Beam test data collected from the Jefferson Lab calorimeter prototype equipped with SiPM readout.}
\label{fig8}
\end{figure}

An example of the energy response of the calorimeter instrumented with SiPMs to PS leptons is shown in Fig.~\ref{fig8}a. The detector was aligned so that the leptons passed through the middle row of the array. To evaluate the energy resolution, we selected leptons that penetrated the central cell of the detector, which appears at the center of the plot.
The PS tile at the center of the plateau corresponds to a lepton energy of approximately 5 GeV. The energy deposited in the central module of the prototype, as well as the sum of energies in all nine detector modules comprising the shower, are shown in Fig.~\ref{fig8}b. Both distributions were fitted using a Crystal Ball function. As seen in the plot, the relative energy resolution, defined as the standard deviation of the distribution divided by the mean, is significantly improved when the shower is reconstructed using all nine prototype modules.

\section{Summary}
\label{sec:misc}

A beam of leptons provided by the Pair Spectrometer in Hall D at Jefferson Lab was used to test various calorimeter prototypes and to perform quality checks on more than one hundred PbWO$_4$ scintillating crystals, which were subsequently used in several calorimeter projects.
The first PbWO$_4$ prototype, instrumented with conventional PMTs, was built and tested in 2018. This was followed by the evaluation of various detectors utilizing SiPM-based readout, including a prototype constructed at Jefferson Lab in 2022 and detectors developed by the CRYTUR company during the same year. The energy resolutions of the SiPM prototypes were found to be slightly worse than those of the PMT version, likely due to differences in the photosensors' active areas. Threshold effects and the linearity of the SiPM matrix were also studied. Since 2020, the SciGlass radiator has been characterized, with multiple prototypes developed and tested.

\acknowledgments        
 
 This work was supported by the Department of Energy, USA. Jefferson Science Associates, LLC operated Thomas Jefferson National Accelerator Facility for the United States Department of Energy under contract DE-AC05-06OR23177.
 
 We greatly appreciate all the Jefferson Lab Physics Division groups and the members of the participating universities for their valuable assistance with prototype fabrication and detector installation in the experimental hall.

\bibliography{report} 
\bibliographystyle{spiebib} 

\end{document}